# Stubby: A Transformation-based Optimizer for MapReduce Workflows[*]

Harold Lim
Duke University
harold@cs.duke.edu

Herodotos Herodotou
Duke University
hero@cs.duke.edu

Shivnath Babu
Duke University
shivnath@cs.duke.edu

## ABSTRACT

There is a growing trend of performing analysis on large datasets using workflows composed of MapReduce jobs connected through producer-consumer relationships based on data. This trend has spurred the development of a number of interfaces—ranging from program-based to query-based interfaces—for generating MapReduce workflows. Studies have shown that the gap in performance can be quite large between optimized and unoptimized workflows. However, automatic cost-based optimization of MapReduce workflows remains a challenge due to the multitude of interfaces, large size of the execution plan space, and the frequent unavailability of all types of information needed for optimization.

We introduce a comprehensive plan space for MapReduce workflows generated by popular workflow generators. We then propose *Stubby*, a cost-based optimizer that searches selectively through the subspace of the full plan space that can be enumerated correctly and costed based on the information available in any given setting. Stubby enumerates the plan space based on plan-to-plan transformations and an efficient search algorithm. Stubby is designed to be extensible to new interfaces and new types of optimizations, which is a desirable feature given how rapidly MapReduce systems are evolving. Stubby's efficiency and effectiveness have been evaluated using representative workflows from many domains.

## 1. INTRODUCTION

Web clicks, social media, scientific experiments, and datacenter monitoring are among sources that generate large quantities of data every day. Rapid innovation and improvements in productivity necessitate timely and cost-effective analysis of this data. This trend is fueling a massive increase in workloads composed of workflows of data-parallel jobs. The jobs are connected with each other through producer-consumer relationships specified by the workflow. MapReduce systems like *Hadoop* [7] and Google MapReduce [4] are now popular choices to run these workflows.

Automatic optimization of these MapReduce workflows is important as well as challenging. The use of data-intensive workflows is growing beyond large Web companies to those with few

[*]Supported by NSF grants 0964560 and 0644106

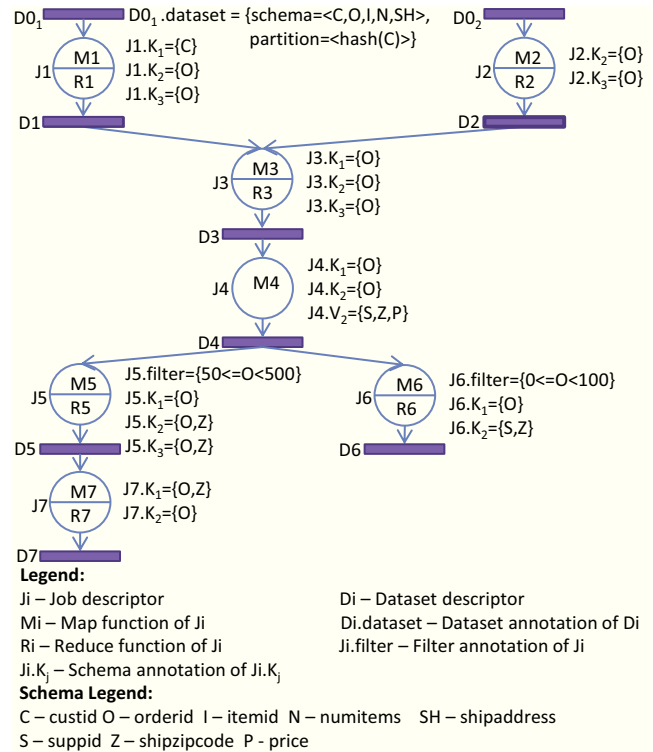

**Figure 1: An example MapReduce job workflow and its annotations (known information) given to Stubby for optimization.**

MapReduce tuning experts. Furthermore, with MapReduce systems being relatively young and evolving rapidly, it is hard to find experienced programmers and administrators to develop and run efficient MapReduce workflows. Recent studies show the order of magnitude performance gap that exists between optimized and unoptimized versions of MapReduce workflows [8, 23].

As an example, consider the MapReduce workflow shown in Figure 1 which is derived from a realistic business report generation application. (We will use this workflow as a running example.) It was convenient for the developer to express the report generation application as a workflow of seven MapReduce jobs. Optimization techniques that we introduce in this paper can automatically convert this seven-job workflow into an equivalent, but highly optimized, two-job workflow. The performance gains are quite dramatic.

The central contribution of this paper is an automatic cost-based optimizer, called *Stubby*[1], for MapReduce workflows. Stubby considers multiple optimization types that can be composed together,

---

[1]The name Stubby (meaning short and stocky) comes from the fact that our workflow optimizer makes workflows shorter (pack-

generating a large plan space for a MapReduce workflow $W$. One optimization type is called *vertical packing* where map and reduce functions from jobs in producer-consumer relationships in $W$ are combined. Vertical packing produces new jobs that avoid the local and network I/O due to shuffling of data between the map and reduce phases of MapReduce execution [4]. For example, vertical packing can be applied to the jobs J5 and J7 in Figure 1, replacing these two jobs with a single job whose reduce function is a combination of J5's reduce function R5, J7's map function M7, and J7's reduce function R7.

Another optimization type is called *horizontal packing* which combines map and reduce functions so that jobs processing the same (large) dataset $d$ can share the read I/O incurred for $d$ [5, 13, 22]. Other optimization types include choices for the partition function of MapReduce jobs, data layouts (e.g., partitioning and compression) of intermediate data read and written by MapReduce jobs, the degree of parallelism to use while running map (reduce) functions as concurrent map (reduce) *tasks*, and many others.

Developing a cost-based optimizer for practical MapReduce workflows poses three nontrivial challenges which we respectively refer to as the *plan spectrum*, *interface spectrum*, and *information spectrum*. The plan spectrum refers to the large and high-dimensional space of possible plans to run a given workflow.

The interface spectrum refers to the many possible ways in which a MapReduce workflow $W$ can be generated in practice. A user could have generated $W$ by writing the map and reduce functions in some programming language for each job in $W$. $W$ could have been generated by query-based interfaces like *Pig* or *Hive* that convert queries specified in some higher-level language to a MapReduce workflow. $W$ could have been generated by program-based interfaces like *Cascading* or *FlumeJava* that integrate workflow definitions into popular programming languages [1].

Furthermore, $W$ could have been generated by composing multiple smaller workflows developed independently [15]. For example, it is natural to generate the workflow in Figure 1 by composing two individual workflows. One component workflow comprises jobs J1, J2, and J3, possibly written in Java, for cleaning and transforming data snapshots taken periodically from OLTP databases. The second component workflow comprises jobs J4, J5, J6, and J7 that are generated from a Pig query that computes various aggregates for report generation. Tools like *Oozie* and *Amazon Elastic MapReduce Job Flow* provide interfaces for such flexible development of MapReduce workflows [15].

The information spectrum refers to a problem endemic to MapReduce systems: the information needed to enumerate or to cost alternate plans considered by an optimization type may not always be available. For example, it is common in MapReduce systems to interpret data (lazily) at processing time, rather than (eagerly) at loading time. Hence, properties of the workflow's input data (e.g., schema, partitioning) may not be known. Lack of such information can make some vertical packing optimizations inapplicable because their correctness cannot be guaranteed. It is also common to have MapReduce programs or user-defined functions written in languages like Java, Python, and Ruby; effectively requiring the optimizer to deal with black-box jobs in workflows. Statistics such as selectivity estimates or processing costs could also be unavailable.

## 1.1 Contributions and Roadmap

Stubby has been designed to address the challenges posed by the plan, interface, and information spectrums. Figure 2 shows how Stubby fits in a MapReduce system. Different interfaces can be ing multiple jobs into one job to reduce workflow height) and fatter (packing multiple parallel function pipelines into one job).

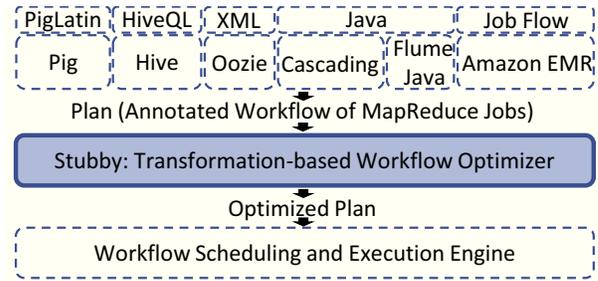

**Figure 2: Stubby in the MapReduce execution stack.**

used to generate the MapReduce workflow given to Stubby for optimization.

Stubby accepts input in the form of an *annotated MapReduce workflow*—which we call a *plan*—and returns an equivalent, but optimized, plan. Annotations are a generic mechanism for workflow generators to convey useful information found during workflow generation. Stubby will find the best plan subject to the given annotations, while working correctly (but not optimally) when zero to few annotations are given. Stubby is also compatible with optimizations that the workflow generator may do, e.g., projection pushdown or join ordering [5, 14, 23].

We designed Stubby as a transformation-based optimizer. A *transformation* is defined by a set of preconditions and postconditions: If the preconditions hold on a plan $P^-$, then Stubby can generate a plan $P^+$ on which the postconditions hold such that $P^-$ and $P^+$ will produce the same result. However, $P^-$ and $P^+$ may have different estimated costs and actual performance. The set of conditions—where a condition may refer to one or more annotations—is a succinct way of capturing the information needed for each optimization type. The combination of transformations and annotations gives Stubby some attractive features to deal with the information and interface spectrums:

- Stubby can search selectively through the subspace of the full plan space that can be enumerated correctly and costed based on the information available in any given setting.
- Stubby's core optimizer-level components for plan enumeration, search, and costing are reusable across different interfaces used to generate MapReduce workflows. Adding a new interface mainly requires writing a component to generate the respective annotations for workflows coming from that interface.
- Similar to extensible optimizers like *EXODUS* [6] developed for database systems, Stubby allows new transformations to be added to extend the optimizer's functionality easily.

The current set of transformations supported by Stubby is described in Section 3. Section 4 will then discuss how Stubby addresses the plan spectrum challenge through a novel enumeration and search algorithm. Section 5 describes how plan costs are estimated. Stubby has been prototyped fully and Section 7 describes a comprehensive evaluation. Notably, we compare Stubby with a baseline that represents how an industrial-strength system (Pig) is used in production today. Stubby consistently outperforms the baseline by 2-4.5X.

## 2. OVERVIEW
## 2.1 MapReduce Workflows

A MapReduce workflow $W$ is a *Directed Acyclic Graph (DAG)* $G_W$ that represents a set of MapReduce jobs and their producer-consumer relationships. Each vertex in $G_W$ is either a MapReduce job $J$ or a dataset $D$. Each edge in $G_W$ is between a job (vertex) $J$ and a dataset (vertex) $D$, and denotes whether $D$ is an input or an output dataset of $J$.

Each MapReduce job $J$ in $G_W$ is of the form $J = \langle p, c, a \rangle$. Here, $p$ represents the MapReduce program that is run as part of $J$. Con-

*figuration* $c$ controls how the program $p$ will be executed as tasks during $J$'s execution [8]. Details of the configuration are given in Section 3.5. Annotations $a$ give any available information about the operation and execution of the program that is relevant for workflow optimization. Annotations are discussed in Section 2.2.

Each dataset $D$ in $G_W$ is of the form $D = \langle d, l, a \rangle$. Here, $d$ represents the dataset's descriptor in the distributed file-system that forms the persistent storage layer of a MapReduce system. *Layout* $l$ controls how $D$ is laid out in the distributed file-system, including how the dataset is partitioned and/or compressed. Stubby currently has support for horizontal partitioning only. The annotations $a$ in this case give any available information about $D$.

**MapReduce Program:** For the purposes of this paper, a MapReduce program is specified by the following four functions [4].[2] All functions except map are optional. $K_1$-$K_3$ and $V_1$-$V_3$ are the respective key and value types.

- *map* function: $map(K_1, V_1) \Rightarrow list(K_2, V_2)$. A map function invocation is made for every key-value pair $\langle K_1 = k_1, V_1 = v_1 \rangle$ in the input dataset. During job execution, the key-value pairs in the input are processed in parallel by a set of *map tasks*. The number of map tasks is determined by the job configuration [8].

- *reduce* function: $reduce(K_2, list(V_2)) \Rightarrow list(K_3, V_3)$. For each unique key $K_2 = k$ in the map output key-value pairs, a reduce function invocation is made with the group of all values that have key $K_2 = k$. The number of reduce tasks is determined by the job configuration [8].

- *combine* function: $combine(K_2, list(V_2)) \Rightarrow list(K_2, V_2)$. For any key $K_2 = k$ in the map output key-value pairs, a combine function may optionally be invoked with two or more values associated with $k$. This function is used by map tasks to preaggregate map outputs to reduce I/O and network costs at the expense of additional compute cost. The invocation of this function can be turned on or off, and its granularity of invocation adjusted, by the job configuration [8].

- *partition* function: $partition(K_2) \Rightarrow partition\ descriptor$. This function is used to partition the map output key-value pairs among the reduce tasks. The default is hash partitioning on key $K_2$ along with sorting the map output key-value pairs on $K_2$ per partition so that pairs with the same value of $K_2$ are grouped together for each $reduce(K_2, list(V_2))$ function invocation. Range partitioning is an alternative to hash partitioning.

## 2.2 Annotations

Annotation is the medium used in Stubby to represent and communicate information needed for the different optimization types applicable to a workflow $W$. Broadly speaking, annotations can be categorized based on whether they represent information about the (i) datasets in $W$, (ii) operations performed by the MapReduce programs in $W$, or (iii) the run-time execution of the programs in $W$. We will next describe the specific annotation types supported currently by Stubby under these three categories. Section 6 will describe how the annotations are generated.

**Annotations for datasets:** *Dataset annotations* expose information known about the datasets in a workflow. Physical design information is the most relevant and includes any known partitioning, ordering, compression, and file-level information for the data as stored on the distributed file-system. For example, the dataset annotation for the base dataset $D0_1$ in Figure 1 conveys to Stubby that $D0_1$ is hash partitioned on an attribute named "custid".

---

[2]MapReduce implementations like Hadoop allow other functions to be specified, e.g., for parsing/splitting map inputs and secondary sorting of map outputs. Our implementation of Stubby for Hadoop supports these additional complexities to a fair extent. We omit the details in order to focus on the research contributions. For ease of exposition and without loss of generality, for any producer job $J_p$ whose output is read by a consumer job $J_c$, we will assume that the key-value pairs output by $J_p$ are input as is to $J_c$'s map function.

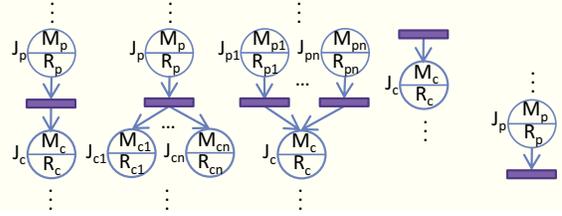

**Figure 3: Five types of producer-consumer subgraphs that can arise in a workflow DAG (some combinations of these subgraphs can also arise).**

**Annotations for programs:** Stubby currently supports two types of annotations—*schema* and *filter*—to expose known properties of MapReduce programs that are otherwise black-boxes to Stubby. Schema annotations expose the composition of the key and value types—$K_1$-$K_3$ and $V_1$-$V_3$—in a MapReduce program. For example, a schema annotation in Figure 1 specifies key $K_2$ in job J5 as consisting of two fields: "orderid" and "shipzipcode". In addition, key $K_2$ in job J7 is the single field "orderid". Identical field names are used in schema annotations to indicate data that flows unchanged through different functions in MapReduce programs. This concept is defined formally in Section 3.1. Schema annotations can be accompanied by filter annotations to convey that a program uses as input only a subset of the dataset generated by its producer job in the workflow (e.g., see jobs J5 and J6 in Figure 1).

**Annotations for program execution:** *Profile annotations* expose statistical information about the run-time execution of a program. This information is useful to estimate the cost of running a program under different data layouts and job configurations. Based on our previous work on the *Starfish* system, we chose to expose two categories of information through profile annotations [8]: (i) Dataflow statistics capture the distribution of key-value pairs and bytes flowing through different phases of a MapReduce program execution; (ii) Cost statistics capture the distribution of execution time spent in different phases of a MapReduce program execution.

## 2.3 Problem Definition and Solution Approach

Given an initial plan $P$ for a MapReduce workflow $W$—namely, the workflow DAG $G_W$ and a set of annotations associated with $W$—the goal of Stubby is to automatically find a plan $P_{opt}$ for $W$ with minimum overall estimated execution cost. The space of possible plans for $W$ is defined by transformations that can be applied to a plan. We categorize these transformations into: (i) intrajob vertical packing transformation, (ii) inter-job vertical packing transformation, (iii) horizontal packing transformation, (iv) partition function transformation, and (v) configuration transformation. Section 3 describes each transformation in terms of its preconditions, postconditions, and required annotations. Sections 4 and 5 describe Stubby's enumeration and search as well as plan costing techniques respectively.

For describing the transformations, we identify five subgraphs that characterize different types of producer-consumer relationships arising among jobs in the workflow DAG. These producer-consumer subgraphs are shown in Figure 3: *one-to-one*, *one-to-many*, *many-to-one*, *none-to-one*, and *one-to-none*.

# 3. TRANSFORMATIONS THAT DEFINE THE PLAN SPACE

## 3.1 Intra-job Vertical Packing Transformation

An intra-job vertical packing transformation converts a MapReduce job into a Map-only job. Suppose M and R respectively denote the map and reduce functions of the job. Without the vertical packing transformation, M will be invoked in the job's map tasks, and R will be invoked in the job's reduce tasks. After the transformation, the M and R functions will be pipelined together and invoked in the new job's map tasks. The data output by M will now be provided directly to R without going through the partition, sort, and shuffle phases of MapReduce job execution.

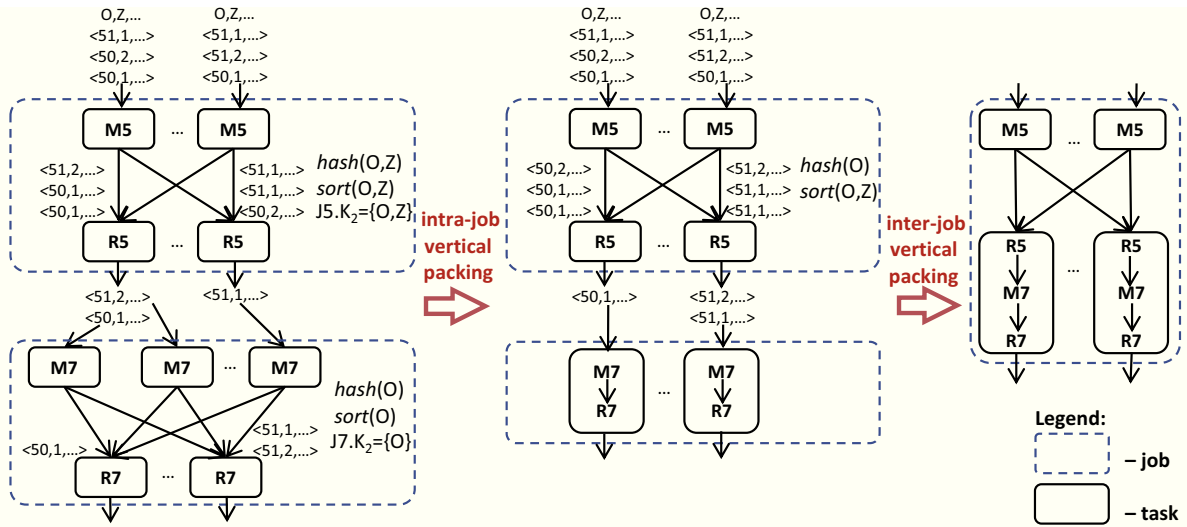

**Figure 4: A task-level illustration of vertical packing transformations applied to the example workflow from Figure 1.**

We will begin with an example of the transformation applied to our example MapReduce workflow from Figure 1. We will then specify formally the preconditions and postconditions for a common case where the transformation applies. This specification will be followed by a discussion of extended scenarios where Stubby will apply the transformation as well as a discussion of the performance implications. A similar presentation style will be used for all other transformations.

Figure 4 shows a task-level view of the one-to-one producer-consumer subgraph comprising jobs J5 and J7 from Figure 1. The plans shown respectively on the left hand side (denoted $P^-$) and the middle (denoted $P^+$) of Figure 4 are the plans before and after applying an intra-job vertical packing transformation to Job J7. Job J5's reduce function R5 needs its input key-value pairs grouped on J5.$K_2$={$O, Z$}, and Job J7's reduce function R7 needs its input grouped on J7.$K_2$={$O$}. As shown on the left side of Figure 4, plan $P^-$ generates both groupings using MapReduce's default strategy: do hash partitioning of the respective map-output key-value pairs on $K_2$, and sort the pairs within each partition on $K_2$.

Plan $P^+$, on the other hand, generates the grouping needed in the producer job J5 differently: a hash partitioning is done on {$O$}, and a per-partition sort is done on the {$O, Z$} combination. The nice property of this grouping technique is that it satisfies the grouping needs of both the producer job J5 and the consumer job J7. Consequently, there is no need to have the partition, sort, and shuffle phases in J7. J7's reduce function R7 can be moved to the map-side and invoked in the map tasks; as shown in plan $P^+$ in Figure 4. Effectively, $P^+$ is pipelining key-values pairs from M7 to R7.

**Preconditions and Postconditions:** Let us build on the intuition from the above example to formalize the preconditions and postconditions for the intra-job vertical packing transformation.[3] Recall that if the preconditions hold on a plan $P^-$, then we can generate a plan $P^+$ on which the postconditions will hold such that $P^-$ and

$P^+$ will produce the same result. However, $P^-$ and $P^+$ may have different performance. We will first consider one-to-one subgraphs and then present extensions.

Preconditions on plan $P^-$ in intra-job vertical packing:

1. There is a one-to-one producer-consumer subgraph with producer job $J_p$ and consumer job $J_c$.

2. The output key-value pairs of the map function $M_c$ of $J_c$ satisfy the following invariant: $M_c$ can output a key-value pair with $J_c.K_2$=k only from one or more key-value pairs with $J_c.K_2$=k given as input to the reduce function $R_p$ of $J_p$. These functions could, in turn, be pipelines of map, reduce, and combine functions due to previous applications of transformations.

Intuitively, the above conditions state that the data in the $J_c.K_2$ fields flows unchanged—allowing for filtering as well as addition or removal of duplicates—from the input of the producer job $J_p$'s reduce function to the output of the consumer job $J_c$'s map function. Stubby checks these conditions based on the schema annotations given in the workflow.

Postconditions on plan $P^+$ in intra-job vertical packing:

1. The partition function of $J_p$ in the new $P^+$ will partition on {$J_p.K_2 \cap J_c.K_2$} and sorts per partition on the combined sort key {$J_p.K_2 \cap J_c.K_2, (J_p.K_2 \cup J_c.K_2) - (J_p.K_2 \cap J_c.K_2)$}; which allows the partition function of $J_p$ to satisfy the reduce-side grouping requirements of both $J_p$ and $J_c$.

2. For any reduce task in job $J_p$, all key-value pairs output by that reduce task should be input in the same order to a single map task in job $J_c$. This requirement can be enforced by specifying a condition on the configuration (recall Section 2.1) of job $J_c$. Note that the map tasks in plan $P^-$ are free to process subsets of key-value pairs output by one or more reduce tasks in job $J_p$.

**Extensions:** With some adjustments, the preconditions and postconditions given earlier for one-to-one subgraphs become applicable to none-to-one and many-to-one subgraphs and their hybrid combinations. For a none-to-one producer-consumer subgraph (e.g., at job J2 in Figure 1), the first postcondition effectively becomes a precondition that should hold on the job's input dataset. Recall that dataset annotations give the partitioning and ordering information required to check whether such conditions hold.

For a many-to-one subgraph (e.g., at job J3 in Figure 1), the second precondition should hold for each producer-consumer pair. The postconditions also need to be adjusted to have the same partitioning on $J_{p_i}.K_2$ for all producer jobs $J_{p_i}$ so that all key-value

---

[3]A proof of the correctness of these conditions is given in the online technical report [12].

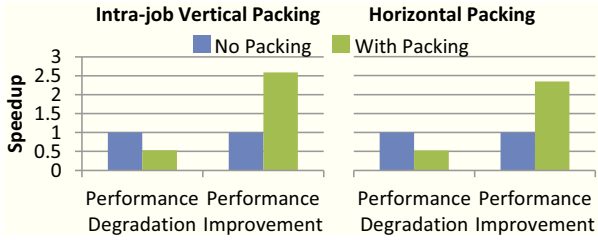

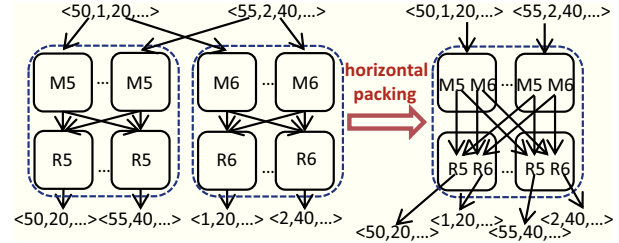

**Figure 5: Performance degradation and improvement caused by vertical packing and horizontal packing transformations.**

**Figure 6: A task-level illustration of horizontal packing applied on jobs J5 and J6 of the example workflow (refer to Figure 1).**

pairs with $J_c.K_2=k$ can be input to a single map task in the consumer job $J_c$.

**Performance Implications:** The new plan $P^+$ produced by an intra-job vertical packing transformation can perform better or worse than the old plan $P^-$; motivating a cost-based approach to decide whether to apply the transformation or not. For illustration, Figure 5 shows the actual performance with and without vertical packing for a none-to-one subgraph when we vary the properties of the input dataset. A 10-node Hadoop cluster on Amazon EC2 is used. (Further details of the experimental setup are given in Section 7.)

Figure 5 shows the speedup over the case of not applying the transformation. Note that, in one case, vertical packing leads to a 2.5X speedup. As expected, the performance gains from applying intra-job vertical packing come from eliminating the large overhead of moving the map output data to the reduce tasks: CPU cost for partitioning and sorting the data, I/O from writing and reading to local disk, as well as network transfer costs.

However, in the other case, vertical packing leads to a 0.5X degradation in performance. Interestingly, there are a number of negative performance effects of vertical packing:

- A vertical packing creates a dependence between the configuration choices for the producer job $J_p$ and consumer job $J_c$, reducing the degrees of freedom in choosing the best plan. The degree of map-side parallelism in $J_c$ is now dependent on the reduce-side parallelism in $J_p$ due to the second postcondition.

- Note that, for job J5 in Figure 4, the application of intra-job vertical packing led to a choice of partitioning on $\{O\}$ in $P^+$, whereas $P^-$ partitions on the $\{O, Z\}$ combination. It is possible that attribute $\{O\}$ has few unique values in the data—one in the worst case—but the $\{O, Z\}$ combination has many unique values. In this case, vertical packing can lead to significant performance degradation by limiting the parallelism in $P^+$.

- In popular MapReduce implementations like Hadoop, map and reduce tasks are run in *task slots* that usually have preconfigured resources (e.g., heap memory). Thus, packing more functions to run in the same task has the potential to cause suboptimal resource usage in one of two ways: (i) resource contention from executing more functions per task slot, and (ii) resource underutilization from using fewer task slots than what is available.

These issues have to be taken into account during plan costing in order to ensure that vertical packing is considered in a comprehensive cost-based fashion.

### 3.2 Inter-job Vertical Packing Transformation

An inter-job vertical packing transformation moves functions from a job $J$ into another job, completely eliminating the need for $J$. The example workflow in Figure 1 shows multiple opportunities for this transformation. For example, since J4's map function M4 is invoked for every key-value pair output by job J3, and does not require any grouping, M4 can be pipelined after J3's reduce function; eliminating reads and writes for the dataset D3. Moreover, a previously-transformed job can be further transformed as shown on the right side of Figure 4.

**Preconditions and Postconditions:** Under the following preconditions, the functions in a Map-only job can be moved to another job as part of an inter-job vertical packing transformation:

1. There is a one-to-one producer-consumer subgraph with (only) one producer job $J_p$ and (only) one consumer job $J_c$.

2. One of $J_p$ or $J_c$ is a Map-only job.

**Extensions:** Multiple choices exist to apply this transformation to a one-to-many producer-consumer subgraph. For example, consider a Map-only producer job $J_p$: (i) The functions of $J_p$ can be replicated and packed with the functions in the map task of each consumer job; or (ii) $J_p$ and one of the consumer jobs can be packed into a single job, while ensuring that $J_p$'s original output dataset is still generated (materialized to disk) for the other consumer jobs.

**Performance Implications:** Similar to intra-job vertical packing, this transformation can have positive or negative performance implications. The performance gains from applying inter-job vertical packing come from eliminating disk and network I/O as well as the overhead of setting up and cleaning up additional map tasks. However, most negative performance effects of intra-job vertical packing apply here as well. If one of the MapReduce jobs has to be run as a single task (e.g., a top-K computation), then an inter-job vertical packing transformation can cause the entire computation to run as a single task; giving extremely poor performance.

### 3.3 Horizontal Packing Transformation

A horizontal packing transformation packs the map (reduce) functions of multiple jobs that read the same dataset into the same map (reduce) task of a transformed job. Jobs J5 and J6 of the example workflow in Figure 1 read the same dataset D4. Figure 6 shows a task-level view of packing jobs J5 and J6 into a single job.

While vertical packing transformations pipeline functions sequentially, a horizontal packing transformation puts multiple map (reduce) functions from separate parallel pipelines into a single job's map (reduce) task. An input key-value pair $\langle K_1, V_1 \rangle$ will go through all pipelines in the map task, and each pipeline will generate its own $\langle K_2, list(V_2) \rangle$ outputs. In the reduce task, each $\langle K_2, V_2 \rangle$ pair will only go through the pipeline that corresponds to the map function that generated the pair.

**Preconditions and Postconditions:** The easy precondition for applying a horizontal packing transformation is that two or more jobs should have the same input dataset, e.g., in a one-to-many producer-consumer subgraph [5, 13].

**Extensions:** The precondition of reading the same input dataset can be relaxed so that a horizontal packing transformation can be applied to any set of concurrently-runnable jobs, e.g., jobs J1 and J2 in our example workflow. The only additional requirement is to ensure that the map functions in separate parallel pipelines only process key-value pairs from the respective input datasets of these functions. In conjunction with the vertical packing transformations, such an extended horizontal packing transformation can transform jobs J1, J2, and J3 of our example workflow into a single job.

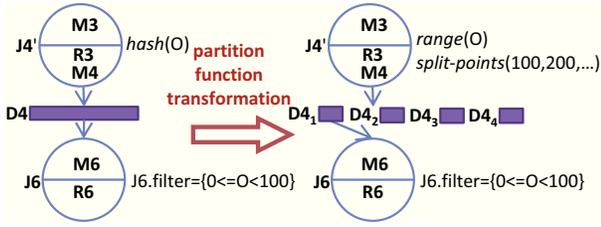

**Figure 7: An illustration of partition function transformation applied on job J4' that transforms the partition function to range partitioning, which enables partition pruning on job J6.**

**Performance Implications:** Figure 5 shows that horizontal packing transformations can lead either to performance gain or to performance degradation. Both experimental results are from a 10-node Hadoop cluster on Amazon EC2. The workflow used has two consumer jobs that perform filtering, grouping, and aggregation on an input dataset. A very large input dataset is used in one case and a smaller dataset in the other.

On the positive side, horizontal packing transformations can improve performance by eliminating local-disk and network I/O from reading the input dataset multiple times. On the negative side:

- A horizontally-packed job essentially runs all individual jobs with the same configuration. This dependence can cause performance issues. For example, the performance degradation for the smaller dataset in Figure 5 was because the cluster had enough resources to run all consumer jobs concurrently and most efficiently; resulting in better performance than when running a single horizontally-packed job. Furthermore, packing multiple functions in parallel per task can cause issues such as excessive spilling of key-value pairs to local disk due to the concurrent memory overheads [5].

- Depending on the selectivity of the map functions, the extra overhead in the packed job from partitioning and sorting the combined map-output data from all individual jobs may outweigh the performance gains from read sharing [13].

### 3.4 Partition Function Transformation

Partition function transformation changes how the map output key-value pairs are partitioned and sorted during the execution of a job. This transformation includes, but is not limited to: (i) changing the partitioning type (default is hash), (ii) changing the splitting points for range partitioning, and (iii) changing the fields on which per-partition sorting happens (default is $K_2$). For example, in Figure 7, this transformation changes the partition function of job J4' from using hash partitioning to range partitioning. (Note that J4' is itself a transformed job that was generated by an inter-job vertical packing of jobs J3 and J4 of the example workflow.)

**Preconditions and Postconditions:** There are no preconditions for a partition function transformation on a job J. The new partition function for J in plan $P^+$ should satisfy all current conditions on the partition function for J in $P^-$. For example, note that the application of an intra-job vertical packing transformation will place a postcondition on the partition function of the producer job. Furthermore, the MapReduce workflow given to Stubby could have some initial conditions already imposed on a job's partition function. For example, a MapReduce job for sorting an input dataset will need to use range partitioning.

**Performance Implications:** Partition function transformation can improve the performance of a job. First, the correct choice of partition function can decrease data skew in the reduce tasks within a single job. When the profile annotation for a job provides the data distribution of map-output key-value pairs, range partitioning with good splitting points can be chosen instead of hash partitioning to ensure that data is distributed evenly across all reduce tasks.

Second, the partition function of a producer job $J_p$ affects the layout of its output dataset. Thus, adjusting the partition function's splitting points based on any filter annotations provided for a consumer job $J_c$ will enable *partition pruning* in $J_c$. With partition pruning, $J_c$ will only read the partitions of $J_p$'s output dataset that are relevant to $J_c$; saving on local and network I/O.

For example, consider job J6 in our example workflow (see Figure 7). J6 discards all input key-value pairs with orderid $\geq 100$ (exposed through the filter annotation). Thus, the partition function of J4' can be transformed to range partitioning (e.g., in ranges of 100) so that J6's input data descriptor can be set to be the partition(s) containing the output of J4' with $0 \leq$ orderid $< 100$.

### 3.5 Configuration Transformation

A configuration transformation changes the configuration of a MapReduce job in a workflow. Figure 8 shows an example of this transformation applied on job J5. Here, J5 is transformed to use 80 reduce tasks, a map output buffer size (for two-phase sorting) of 512 MB, and compression is turned on for the map and reduce output key-value pairs (in turn, affecting dataset D5's layout).

**Preconditions and Postconditions:** There are no preconditions for a configuration transformation on a job J. The new configuration for J in plan $P^+$ should satisfy all current conditions on the configuration for J in $P^-$. For example, recall from Section 3.1 that the application of an intra-job vertical packing transformation will place a condition on the configuration of the consumer job.

**Performance Implications:** As observed in [8], the configuration space for a MapReduce job is large and high-dimensional. In Hadoop, for example, a job's performance is controlled by the settings of dozens of parameters such as those shown in Figure 8. The respective performance impacts of these parameters are correlated and vary based on the properties the MapReduce program, input datasets, and cluster resources. Furthermore, the configuration transformation applied on a producer job J not only affects J's performance, but also the performance of the consumer jobs that read J's output. Thus, nontrivial cost-based decisions have to be made in order to pick the best configurations for jobs in a workflow.

### 4. SEARCH STRATEGY

Given a plan $P$ (i.e., an annotated MapReduce workflow $W$), Stubby's goal is to find the sequence of valid transformations to apply to $P$ in order to generate an equivalent plan $P'$ that minimizes the overall execution time of $W$. Different sequences of transformations can generate very different plans. For example, consider the MapReduce workflow in Figure 1. One option is to apply the intra-job vertical packing transformation on job J7, followed by the inter-job vertical packing transformation, in order to pack jobs J5 and J7 into a single job (as shown in Figure 4). Alternatively, we can apply the horizontal packing transformation on jobs J5 and J6 to generate a different packed job, as shown in Figure 6. The *Plan Space* $S_P$ for plan $P$ consists of all valid alternative plans for $P$ generated by applying combinations of transformations to $P$.

**Workflow Optimization Process:** One approach to optimize a plan $P$ is to apply enumeration and search techniques to the full plan space $S_P$. However, the large size of $S_P$ renders this approach impractical. More efficient search techniques can be developed based on two key insights. The first insight comes from how transformations interact with each other. In theory, a decision to apply any transformation on a particular job in $P$ can influence the choice of a transformation on any other job in the same plan. However, in practice—primarily due to the semantics and implementation of the MapReduce programming model—arbitrary interactions among transformations across multiple jobs are uncommon. Consider again the example workflow from Figure 1. The decision to apply an inter-job vertical packing transformation on jobs J3 and J4 does not affect the transformations that are applicable to job J7; therefore, these decisions can be made independently.

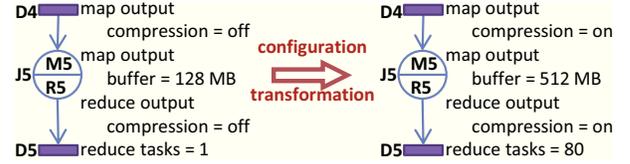

**Figure 8: An illustration of configuration transformation applied on job J5 of the example workflow.**

Thus, we follow a *divide-and-conquer* approach: $P$ is divided into (possibly overlapping) subplans, denoted $P^{(i)}$, with smaller plan subspaces $S_P^{(i)}$ such that the globally-optimal choice in $S_P$ can be found by composing the optimal choices found for each $S_P^{(i)}$. Each $P^{(i)}$, along with the corresponding $S_P^{(i)}$, defines an *Optimization Unit* $U^{(i)}$. The idea behind an optimization unit is to bring together a set of related decisions that affect each other, but are independent of the decisions made at other optimization units. In other words, the goal is to break the large plan space $S_P$ into independent subspaces $S_P^{(i)}$ such that $S_P = \cup S_P^{(i)}$. Within each $U^{(i)}$, Stubby is responsible for enumerating and evaluating the different transformations applicable to the jobs in $U^{(i)}$.

The second key insight is that the order of applying transformations is important if we prefer to avoid expensive backtracking techniques. Applying a transformation may enable the use of another transformation (e.g., an intra-job vertical packing transformation on job J7 enables an inter-job vertical packing between J5 and the new J7' to eliminate one entire job) or it may prevent it (e.g., a horizontal packing transformation on jobs J5 and J6 prevents an intra-job vertical packing transformation on job J7). Therefore, it is essential to guide the search efficiently towards a sequence of transformations that can lead to near-optimal execution plans.

We organize transformations in two (overlapping) groups. The first group, termed *Vertical*, focuses on applying intra- and inter-job vertical packing transformations. The second group, termed *Horizontal*, focuses on applying the horizontal packing transformation. The aforementioned transformations are unique in the sense that, once applied, they change the structure of the workflow graph. On the other hand, the partition function and configuration transformations do not change the graph structure. These two transformations are included in both the Vertical and the Horizontal groups.

The Vertical transformations are applied within all optimization units before the Horizontal transformations are considered. This ordering stems from two observations. First, for the new horizontally-packed job, the horizontal packing transformation creates a map-output key $K_2$ that combines the $K_2$ keys from the original jobs. This new, and possibly complex, key can prevent the application of vertical packing transformations on succeeding jobs. Following our running example from Figure 1, applying horizontal packing to jobs J5 and J6 will prevent using intra-job vertical packing on job J7 because the preconditions can no longer be met. Second, intra- and inter-job vertical packing transformations can potentially bring higher benefits as they eliminate entire shuffle steps as well as writing and reading intermediate data between jobs. On the other hand, horizontal packing transformations can only reduce the amount of data read through scan sharing.

Overall, Stubby's optimization process is as follows:

**Step 1.** Generate the first optimization unit consisting of one or more jobs in the MapReduce workflow graph $G_W$ (described in Section 4.1).

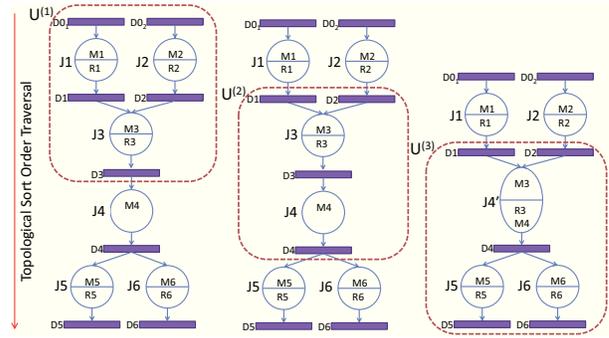

**Figure 9: An illustration of Stubby's dynamic generation of optimization units as it traverses the example workflow graph.**

**Step 2.** Enumerate and search within an optimization unit $U$ using the Vertical transformations in order to find the (near) optimal subplan for $U$ (described in Section 4.2). These transformations may alter the structure of the subgraph in $U$.

**Step 3.** Dynamically generate the next optimization unit in $G_W$ in topological sort order, and apply Step 2.

**Step 4.** Repeat Step 3 until the entire graph $G_W$ is covered.

**Step 5.** Repeat Steps 1-4 using the Horizontal transformations to find the overall (near) optimal execution plan for $W$.

## 4.1 Dynamic Generation of Optimization Units

Stubby builds the optimization units dynamically based on the following observation: when two jobs $J_i$ and $J_k$ are separated by one or more jobs in the workflow graph (i.e., the dependency path between $J_i$ and $J_k$ contains at least one other job), then the effect of $J_i$ on the execution of $J_k$ diminishes rapidly in practical settings. Hence, decisions for $J_i$ can be made independently from decisions made for $J_k$. For example, the choice for applying inter-job vertical transformation on jobs J3 and J4 in our example workflow from Figure 1, will not affect the choice for using an intra-job vertical transformation on job J7.

Each optimization unit $U^{(i)}$ consists of a set of concurrently-runnable producer jobs and the corresponding set of consumer jobs. Figure 9 offers a pictorial representation of the optimization units. The first optimization unit $U^{(1)}$ (denoted by a dotted box in Figure 9) consists of the producer jobs J1 and J2 as well as the consumer job J3. The plan space $S_W^{(1)}$ contains the subplans formed by all valid combinations of transformations that can be applied on jobs J1, J2, and J3.

Applying transformations within an optimization unit may alter the structure of the graph. As an example, suppose only configuration transformations are beneficial to reduce the total running time of the jobs in $U^{(1)}$. In this case, the structure of the graph remains unchanged. Since Stubby traverses the graph in topological sort order, the next optimization unit $U^{(2)}$ will be generated with J3 as the producer job and J4 as the consumer job (see Figure 9). Now suppose that the best transformation to apply is inter-job vertical packing to job J4. This transformation will replace jobs J3 and J4 with a new job J4'. The next optimization unit $U^{(3)}$ will consist of the new producer job J4' and the consumer jobs J5 and J6.

## 4.2 Search Within an Optimization Unit

For each optimization unit $U^{(i)}$, Stubby must find the subplan from $S_W^{(i)}$ that minimizes the total running time of the MapReduce jobs in $U^{(i)}$. Stubby addresses this problem by generating alternative valid subplans using transformations selected through an enumeration and search over $S_W^{(i)}$.

The number of jobs within any individual optimization unit $U^{(i)}$ is typically small. We observed that applying all combinations of transformations apart from the configuration transformation within $U^{(i)}$ usually results in a small number of unique subplans. Therefore, Stubby exhaustively applies all transformations, except the configuration transformation, in order to generate all possible subplans $p_1$–$p_n$ for $U^{(i)}$. For example, as illustrated in Figure 10, this exhaustive enumeration for optimization unit $U^{(2)}$ from Figure 9, generates only four alternative subplans $p_1$–$p_4$.

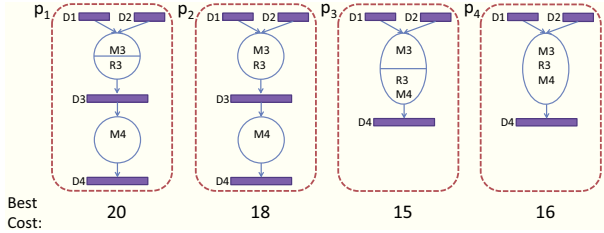

| | $p_1$ | $p_2$ | $p_3$ | $p_4$ |
|---|---|---|---|---|
| Best Cost: | 20 | 18 | 14 | 16 |

**Figure 10: Enumeration of all valid transformations for optimization unit $U^{(2)}$ from Figure 9. The corresponding best estimated cost (running time) from RRS invocation is also shown.**

Configuration transformations are applied on the jobs in each generated subplan $p_i$. These transformations can change any of the numerous MapReduce job configuration parameter settings including the number of map and reduce tasks, memory allocation settings, controls for I/O and network usage, and others [8]. In order to search the large and high-dimensional space of configuration transformations efficiently, Stubby uses *Recursive Random Search (RRS)*. RRS is a fairly recent technique developed to solve blackbox optimization problems [24].

RRS first samples the configuration space randomly in order to identify promising regions that contain the optimal configuration setting with high probability. It then samples recursively in these regions which either move or shrink gradually to locally-optimal settings based on the samples collected. RRS then restarts random sampling in order to find a more promising region to repeat the recursive search. Each transformed subplan generated for $p_i$ through RRS is associated with an estimated cost (see Section 5). The output of RRS for $p_i$ is the configuration transformation that leads to the subplan $p_i^{(opt)}$ with the lowest estimated cost for $p_i$. After RRS has been invoked for all the subplans $p_1$–$p_n$ in the optimization unit $U^{(i)}$, Stubby will select the $p_i^{(opt)}$ with the overall lowest estimated cost as the best subplan for $U^{(i)}$.

Consider the example in Figure 10 which shows the four subplans $p_1$–$p_4$ for the optimization unit $U^{(2)}$ from Figure 9. RRS will be invoked four times for $U^{(2)}$ in order to find the best configuration transformation and estimated cost for each $p_i$. When the RRS invocations complete, Stubby will choose to retain subplan $p_3$ from Figure 10 which has the lowest estimated cost among $p_1$–$p_4$. Note that $p_3$ was generated by applying the inter-job vertical packing transformation on job J4.

**Overall Optimization Process:** In summary, Stubby uses a *two-phase greedy* enumeration and search strategy. In each phase, Stubby generates optimization units dynamically while traversing the workflow graph in topological sort order. In the first phase, the producer jobs in each optimization unit $U^{(i)}$ are optimized by applying transformations from the Vertical group. At the end of the optimization process within $U^{(i)}$, (only) the best subplan for $U^{(i)}$ is retained by applying the corresponding transformations to the jobs in $U^{(i)}$. After the entire graph is traversed once, the above process is repeated once more. However, in this second phase, transformations from the Horizontal group are applied. The fully-optimized workflow graph is ready when the second traversal completes.

## 5. PLAN COSTING

For each annotated MapReduce workflow $W$ that is generated during the enumeration and search strategy described in Section 4, Stubby must estimate the execution cost of $W$. Stubby uses Starfish's *What-if Engine* for this purpose [8]. The Starfish What-if Engine is given four inputs:

1. The dataflow and cost statistics of each job in $W$ (recall the profile annotations discussed in Section 2.2).
2. The configuration to run each job in $W$ with (chosen by RRS).
3. The size and layout information for $W$'s input datasets (recall the dataflow annotations discussed in Section 2.2).
4. The cluster setup and resource allocation that will be used to run $W$. This information includes the number of nodes and network topology of the cluster, the number of map and reduce task slots per node, and the memory available for each task execution.

The Starfish What-if Engine uses these inputs and a mix of analytical, black-box, and simulation models to reason about the impact of configuration settings, data properties, and cluster resource properties on the execution of each MapReduce job $J$ in $W$. The What-if Engine will then output cost estimates for each job as well as the entire workflow. Because of space constraints, we refer the reader to [8] for a detailed description of the Starfish What-if Engine. If any of the inputs required to use the What-if Engine are unavailable—e.g., profile or dataset annotations are not provided in the workflow—then the cost estimation will have to fall back to a simpler cost model such as the number of jobs as used in [11].

One challenge while using the Starfish What-if Engine is that Stubby's vertical and horizontal packing transformations change the jobs in $W$. For example, the intra-job vertical packing transformation will change a MapReduce job into a Map-only job. Thus, the packing transformations have to generate new annotations—in particular, the dataflow and cost statistics—for the new jobs that they generate. This process is called *adjustment* in Stubby since the new annotations are generated by modifying the old ones.

Space constraints preclude the discussion of all adjustments. The adjustments that Stubby uses for profile annotations are motivated by cardinality estimation techniques used in database systems. For instance, during an intra-job vertical packing transformation, the reduce function is moved into the map task and is executed after the map function. The new map-task record selectivity[4] is calculated as the product of the record selectivities of the old map and reduce functions. On the other hand, the CPU cost of the new map task is calculated as the sum of the CPU costs of the old functions.

## 6. IMPLEMENTATION

We have implemented Stubby as a standalone system that can be employed by the many interfaces used to generate MapReduce workflows, as shown in Figure 2. To this extent, we have added a new feature in Apache Pig [18] for exporting and importing annotated MapReduce workflows used by Stubby. Pig was only a choice of convenience; our work applies to arbitrary MapReduce workflows.

**Annotations:** As described in Section 3, some transformations in Stubby require additional information which is expressed as annotations. We have made some minor modifications to the compilation process in Pig—which translates a Pig Latin query to a MapReduce workflow—to automatically extract any available schema, filter, and dataset annotations. The details are given in the online technical report [12]. For example, the composition of the key and value types in a MapReduce job are extracted based on any schema information included in the Pig Latin query. Filter annotations are generated based on any filter statements contained in the query. We generate profile annotations using Starfish's *Profiler* which collects profiles through dynamic instrumentation of unmodified MapReduce workflows [8].

---

[4]Record selectivity is defined as the ratio of the number of output key-value pairs over the number of input key-value pairs.

| Abbr. | Workflow | Dataset Size |
|-------|----------|--------------|
| IR | Information Retrieval | 264 GB |
| SN | Social Network Analysis | 267 GB |
| LA | Log Analysis | 500 GB |
| WG | Web Graph Analysis | 255 GB |
| BA | Business Analytics Query | 550 GB |
| BR | Business Report Generation | 530 GB |
| PJ | Post-processing Jobs | 10 GB |
| US | User-defined Logical Splits | 530 GB |

**Table 1: MapReduce workflows and corresponding data sizes.**

**Transformations and Execution:** Recall from Section 3 that vertical packing transformations chain multiple functions together for execution within the same map or reduce task. Similarly, horizontal packing transformations bring multiple independent functions into the same task. These transformations require the use of wrapper MapReduce classes to execute multiple functions inside a map or a reduce task. In addition, horizontal packing needs a tagging mechanism for guiding the data correctly through the different function pipelines. The Pig execution engine already offered support for wrapper classes and tagging, so only minor modifications had to be made in order to execute Stubby-generated plans correctly.

# 7. EXPERIMENTAL EVALUATION

In our experimental evaluation, we used a Hadoop cluster running on 51 Amazon EC2 nodes of the m1.large type. Each node has 7.5 GB memory, 2 virtual cores, 850 GB local storage, and is set to run at most 3 map tasks and 2 reduce tasks concurrently. Thus, the cluster can run at most 150 map tasks in a concurrent map wave, and at most 100 reduce tasks in a concurrent reduce wave.

For evaluation, we selected representative MapReduce workflows from several application domains. These MapReduce workflows are listed in Table 1 and described in detail in Section 7.1. All workflows are expressed in Pig Latin and executed using the Pig execution engine running on Hadoop.

For comparison purposes, we established a *Baseline* that represents how an industrial-strength system (Pig) is used in production today. In particular, we enabled all (rule-based) optimizations supported by Pig and manually-tuned the configuration parameter settings using rules-of-thumb found in [3].

Our evaluation methodology is as follows:

1. We present the overall performance improvements achieved by Stubby, as well as the performance improvements observed when only a subset of the plan space is considered (Section 7.2).

2. We compare the performance benefits from Stubby against other state-of-the-art techniques (Section 7.3).

3. We evaluate the efficiency of Stubby in terms of its overheads while optimizing MapReduce workflows (Section 7.4).

4. We provide a closer look at how Stubby works within an optimization unit to enumerate and find the best transformations (Section 7.5).

## 7.1 MapReduce Workflows

**Information Retrieval:** Term Frequency-Inverse Document Frequency (TF-IDF) is a representative workflow from the information retrieval domain. TF-IDF calculates weights representing the importance of each word to a document in a collection. The TF-IDF weight is a function of the normalized frequency of a word in a document and the number of documents that contain the word. The default TF-IDF workflow consists of three jobs that calculate: (a) the frequency of a word in a document, (b) the total number of words in each document, and (c) the number of documents containing each word as well as the TF-IDF weight of each ⟨word,document⟩ pair. The input dataset is a randomly generated corpus that is partitioned on the document name.

**Social Network Analysis:** A workflow from the social network analysis domain is used to find the top 20 coauthor pairs who have collaborated most frequently with each other. The input dataset is a list of randomly generated ⟨paperID, authorID⟩ pairs from a power-law distribution, partitioned on {paperID}. The workflow consists of four jobs $J_1$–$J_4$: $J_1$ combines all authors for each paper; $J_2$ creates and counts the coauthor pairs; $J_3$ samples the data and creates partition split points for $J_4$; and $J_4$ finds the top 20 coauthor pairs in decreasing order.

**Log Analysis:** Pavlo et. al. [17] describe a complex join task from the log analysis domain. The workflow uses two input datasets: uservisits (partitioned on {date}) and pageranks. We use the data generator provided in [17] to generate the two datasets. This workflow consists of four jobs. The first job filters uservisits by a specified date range and joins it with pageranks on page url. The second job performs an aggregation to find the average pagerank and total ad revenue, grouped by user. The third job samples and creates partition split points for the last job. The last job finds the user with the highest total ad revenue.

**Web Graph Analysis:** PageRank [16] is an example of a web graph analysis algorithm that finds the ranking of web pages based on the hyperlinks pointing to each page. This algorithm can be implemented as an iterative workflow where each iteration is composed of two jobs. The first job joins on the {pageID} key of the two datasets: (a) the adjacency list with each web page and its outgoing hyperlinks, and (b) the current pagerank of each web page. The second job calculates the new pagerank of each web page. We generated an adjacency list of web pages from a power-law distribution.

**Business Analytics Query:** Query 17 from the TPC-H benchmark is a representative example of a complex business analytics (SQL) query [20]. This query determines how much yearly revenue would be lost on average if orders were no longer filled for small quantities of certain parts. Query 17 generates a four-job workflow. Job $J_1$ scans and processes the lineitem table. Job $J_2$ applies a filter condition on the part table, joins the output of $J_1$ and the filtered part table, and finds the average quantity of each part. Job $J_3$ performs another filtered join on the outputs of $J_1$ and $J_2$. The final job $J_4$ calculates the total price of all parts. We use the TPC-H data generator to generate the input datasets for this workflow. The tables lineitem and part are both partitioned on {partID}.

**Business Report Generation:** Business report generation often involves multiple queries (e.g., that perform different groupby aggregates) on a single source dataset [2]. We emulate this scenario by creating a seven-job workflow that processes the lineitem table from the TPC-H Benchmark. The first job scans and performs an initial processing of the data. Two jobs read, filter, and find the sum and maximum of the prices for the {orderID, partID} and {orderID, supplierID} groupings respectively. The results of these two jobs are further processed by separate jobs to find the overall sum and maximum prices for each {orderID}. Finally, the results are used separately to find the number of distinct aggregated prices.

**Post-processing Jobs:** It is common in MapReduce deployments to have workflows that only operate on small datasets (e.g., in the order of GBs). These workflows would only use a small portion of the resources available in the cluster. For example, small datasets can result from filtering or aggregation operations. To capture this scenario, we created a three-job workflow that operates on a small dataset. The first job scans and performs an initial processing of the data. The other two jobs are groupby-aggregates that compute covariance and correlation respectively on the output of the first job. We use the TPC-H data generator to generate the input dataset for this workflow.

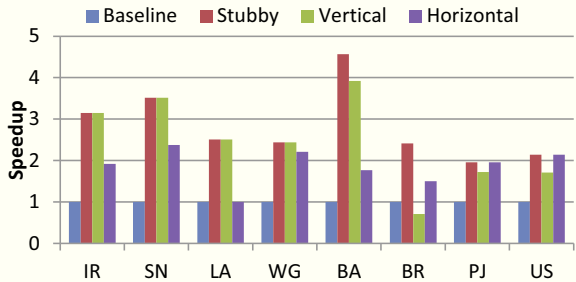

**Figure 11: Speedup over the Baseline achieved by Stubby, Vertical, and Horizontal.**

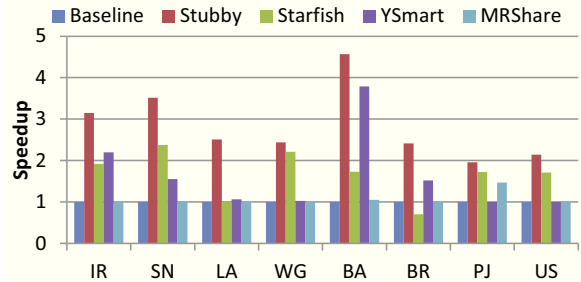

**Figure 12: Speedup over the Baseline achieved by Stubby, Starfish, YSmart, and MRShare.**

**User-defined Logical Splits:** It is common for users to specify logical splits for a set of jobs in a workflow in order to analyze different subsets of data records differently. For example, a Web portal log analysis workflow may want to perform different types of analysis based on specific age groups of users. We created a three-job workflow to emulate this scenario. The workflow consists of a preprocessing (producer) job that outputs the data needed by two consumer jobs. Each consumer job processes only a subset of this data by filtering records in the map function.

### 7.2 Breakdown of Performance Improvements

First, we evaluate the overall improvement given by Stubby on workflow performance. We also evaluate the improvements offered by our two groups of transformations (Vertical and Horizontal) when used in isolation. This breakdown allows us to study the source of improvements obtained from using Stubby. Figure 11 shows the speedup over the Baseline performance achieved by (i) Stubby with all transformations enabled, (ii) Stubby while using only the Vertical group transformations (denoted Vertical), and (iii) Stubby while using only the Horizontal group transformations (denoted Horizontal). Overall, Stubby is able to achieve between 2X and 4.5X speedup over the Baseline. As seen in the figure, the improvements vary depending on the workflow.

For the Information Retrieval (IR), Social Network Analysis (SN), Log Analysis (LA), and Web Graph Analysis (WG) workflows, the performance gains are predominantly due to the vertical packing transformations. These workflows do not present any opportunity for horizontal packing. The speedup achieved by Horizontal is primarily due to the cost-based selection of configuration transformations. The results for these workflows also reflect the spectrum of performance gains we can get from the different packing transformations. For example, Vertical achieves a 2.5X speedup over Horizontal for Log Analysis, whereas the speedup is only 0.2X for Web Graph Analysis. The computation in job $J_2$ of PageRank dominates the overall running time of the workflow, so vertically packing it with job $J_1$ offers limited benefit.

The Business Analytics Query (BA) shows a scenario where both vertical and horizontal packing contribute to the overall performance gains from Stubby. Specifically, the intra-job vertical packing transformation is applicable to the two join jobs in BA (jobs $J_2$ and $J_3$). Since both $J_2$ and $J_3$ process the dataset produced by the first job $J_1$, horizontal packing is also applicable. Stubby applies both transformations to obtain higher benefits compared to using Vertical or Horizontal alone.

The Business Report Generation (BR) workflow is a notable case. Vertical transforms the seven-job workflow into a five-job workflow. However, Vertical performs worse than Horizontal because the nature of BR makes it well suited for benefiting from horizontal packing transformations. Vertical also performs worse than Baseline because we have enabled Pig to use its rule-based optimizations (one of which is horizontal packing). By applying transformations from both the Vertical and Horizontal groups, Stubby generates a three-job workflow that gives a 2.4X speedup.

The Post-processing Jobs (PJ) workflow offers an example where horizontal packing is a wrong decision. Since Baseline performs horizontal packing whenever possible, it generates a suboptimal plan for this workflow. Stubby and Horizontal, being cost-based, correctly decide not to perform horizontal packing for PJ in this case. Furthermore, unlike Baseline, the three other approaches apply the configuration transformation in a cost-based fashion, leading to the performance benefits seen for PJ in Figure 11.

The User-defined Logical Splits (US) workflow is one case where the partition function transformation applies. Specifically, the partition function in the producer job can be changed from the default hash partitioning to range partitioning; thereby enabling partition pruning to be applied to the data read by each consumer job in US.

Overall, it is apparent that different workflows present different transformation opportunities. Stubby is able to recognize and take advantage of these opportunities appropriately to offer speedups ranging from 2X to 4.5X over the Baseline.

### 7.3 Comparison against State-of-the-Art

In this section, we compare Stubby against the following three state-of-the-art approaches for optimizing MapReduce workflows:

1. *Starfish*, based on a cost-based approach proposed in [8], to find good configuration parameter settings for each MapReduce job in the workflow.

2. *YSmart*, based on a rule-based approach proposed in [11], to perform vertical and horizontal packing transformations aggressively in order to minimize the number of jobs in the workflow. We have enhanced YSmart with a rule-based approach for selecting configuration parameter settings.

3. *MRShare*, based on a cost-based approach proposed in [13], to perform horizontal packing transformations. A rule-based approach is used for selecting configuration parameter settings.

Figure 12 shows the speedup achieved over the Baseline after optimizing our eight workflows using Stubby, Starfish, YSmart, and MRShare. Overall, the other approaches are all able to achieve good speedups over the Baseline, with the speedup value depending on the workflow. Stubby is able to outperform all other approaches for all workflows since Stubby considers a strict superset of the optimization opportunities that the others consider, and in a cost-based fashion. For example, Stubby is the only optimizer that considers the opportunity to prune partitions through partition

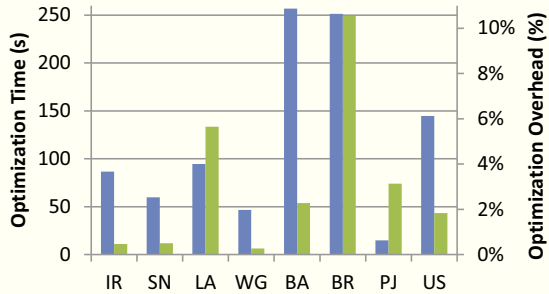

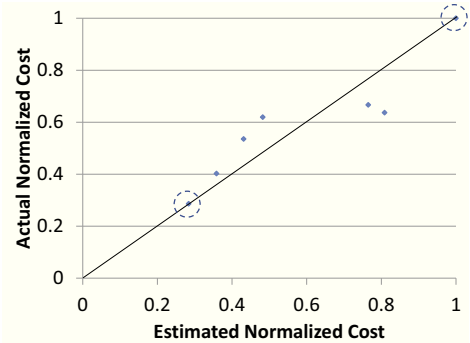

**Figure 13: Optimization overhead for all workflows in terms of (a) absolute time, and (b) a percentage over the total running time of each workflow.**

function selection for the Log Analysis and User-defined Logical Splits workflows.

From the speedups that Starfish achieves in Figure 12 (ranging between 1.5X and 2.4X), we observe that finding good configuration parameter settings in a cost-based fashion can give significant performance improvements. However, Starfish misses out on all vertical and horizontal packing opportunities that can provide significantly higher speedups, like in the case of the Business Analytics Query (BA).

YSmart and MRShare do not automatically find good configuration settings to use. For example, YSmart is able to achieve a 1.5X speedup for the Social Network Analysis (SN) workflow from performing vertical packing. With better configuration settings, Stubby is able to increase the speedup to 3.5X. Similarly, MRShare is able to achieve a 1.4X speedup for the Post-processing Jobs (PJ) workflow, whereas Stubby can achieve close to 2X speedup from selecting better configuration settings.

With a rule-based approach that tries to minimize the number of MapReduce jobs, YSmart can sometimes make suboptimal decisions. This case was evident in the Post-processing Jobs (PJ) workflow where YSmart performed horizontal packing on the two consumer jobs. Stubby and MRShare, on the other hand, used their cost-based approach to determine that horizontal packing was not a good choice, and chose to have the two jobs run independently.

Finally, as MRShare only considers the horizontal packing transformation, it does not provide any performance improvements for many of the MapReduce workflows considered in our evaluation.

### 7.4 Optimization Efficiency

In this section, we evaluate the efficiency of Stubby in finding near-optimal transformations to apply to a given MapReduce workflow. Figure 13 shows the optimization time of Stubby in seconds as well as a percentage over the Baseline running time. Stubby spent on average less than 2 minutes to optimize each workflow. In the worse case, Stubby took around 5 minutes for optimizing the Business Analytics Query (BA) and Business Report Generation (BR) workflows, which contain 4 and 7 jobs respectively.

Percentage-wise, the optimization overhead for seven out of the eight workflows is less than 6%. At worst, Stubby introduced an overhead of 10.5% for the BR workflow which is our largest workflow with 7 jobs. Overall, Stubby's optimization overhead is small compared to the 2X to 4.5X speedup that Stubby gives for these workflows (recall Section 7.2). Since many analytical workflows are run periodically, the optimization overhead of Stubby can be amortized over multiple workflow runs.

### 7.5 Deep Dive into an Optimization Unit

As discussed in Section 4, Stubby (i) enumerates all combinations of valid transformations within an optimization unit in order to generate all valid subplans, (ii) finds the best job configurations for each subplan, and (iii) selects the subplan with the lowest estimated cost. In this experiment, we drill down into the first optimization unit $U^{(1)}$ of the Information Retrieval (IR) workflow. In $U^{(1)}$, seven distinct combinations of transformations can be applied to yield seven subplans $p_1$–$p_7$.

**Figure 14: Actual vs. estimated normalized cost for all combinations of valid transformations in the first optimization unit of the Information Retrieval workflow.**

We captured the best configuration settings generated by Stubby for each subplan $p_i$ and used them to run each $p_i$ separately. Figure 14 shows the scatter plot of the actual and estimated normalized costs for the seven subplans. Ideally, the points in the scatter plot should fall on the solid line. The inaccuracies are due to measurement errors during profiling and estimation errors when calculating plan costs [8]. We observe that the cost estimates are good enough for Stubby to identify the subplans that will lead to the best and worst performance (indicated by dotted circles in Figure 14).

## 8. RELATED WORK

A number of recent projects provide users with various interfaces for generating data-parallel workflows [1, 9, 18, 26]. DryadLINQ and FlumeJava provide libraries and classes for specifying workflows using popular programming languages such as C# and Java respectively [1, 26]. On the other hand, systems like Hive, Pig, Jaql, and SCOPE provide their own high-level declarative languages for creating MapReduce workflows [9, 18, 27]. Our work on Stubby is complementary to these projects in that Stubby is designed to support different interfaces by sitting directly above the workflow scheduling and execution engine (refer to Figure 2). The optimization techniques that we introduce in this paper can be applied to any MapReduce workflow regardless of the interface used to generate the workflow.

There is a large body of work on automatically optimizing workflows of data-parallel jobs [8, 10, 11, 13, 14, 22, 25]. The techniques used can be categorized as either rule-based, such as Flume-Java [1], Manimal [10], and YSmart [11], or cost-based, such as MRShare [13] and Starfish [8]. This category of work differs from Stubby in one or more ways such as: (a) considering a much smaller plan space, (b) focusing on some specific interface, or (c) relying on the guaranteed availability of specific types of information.

YSmart translates SQL-like queries into a set of MapReduce jobs based on four primitive job types: selection-projection, aggregation, join, and sort. YSmart's rule-based optimizer then uses the knowledge of the job primitives used in the queries in order to merge MapReduce jobs. YSmart's goal is to minimize the total number of jobs, which can occasionally lead to suboptimal plans in terms of actual performance. Also, YSmart does not consider optimization opportunities available from partition function transformations and configuration transformations. Similarly, FlumeJava uses information regarding the provided Java class abstractions to pack higher-level operations into the minimum number of MapReduce jobs. MRShare focuses on optimization of multiple MapReduce jobs by applying cost-based decisions for horizontal packing transformations on the jobs. MRShare does not consider workflows or vertical packing. Starfish proposes a cost-based approach for applying (only) configuration transformations.

In contrast, Stubby considers a much larger plan space for workflow optimization that subsumes the plan spaces covered by each of the previously mentioned works. Furthermore, Stubby is designed to be a general-purpose system for workflow optimization where workflows can be optimized regardless of the interfaces used and availability of information. Stubby is able to consider the correct subspace of the full plan space based on the information available. For example, if schema annotations are not available, then Stubby will not consider intra-job vertical packing transformations.

While Stubby considers a large plan space, there are transformations that are not supported by Stubby currently. For example, Wu et al. [23] develop cost-based query optimization techniques for multi-way join queries in MapReduce systems. Their approach automatically translates a user-submitted query into a final plan of MapReduce jobs by optimizing operator selection and ordering for joins. A transformation-based optimizer has been developed for the SCOPE system [27]. The focus of this optimizer is on how partitioning, sorting, and grouping properties can be exploited to avoid unnecessary operations during parallel processing of relational operators. FTOpt [21] introduces the space of fault-tolerance plans for workflows and then uses a cost-based approach to select the best fault-tolerance strategy for each job of a workflow.

The vertical packing transformations in Stubby are related to work on optimizing the computation of multiple aggregates over the same or similar sets of grouping attributes (e.g., [2]). Stubby is also related to work done on optimizing workflows of extract-transform-load (ETL) processes and business processes. For example, Simitsis et al. converted the problem of optimizing ETL workflows into a state space search problem where each state is a graph representation of the workflow [19]. The authors introduced rules for generating equivalent states and used a greedy heuristic search algorithm to find the optimal state.

## 9. CONCLUSIONS

As the popularity of MapReduce for big data analytics grows, the software ecosystem around MapReduce is also growing rapidly to provide users with different interfaces for generating MapReduce workflows. However, automatic cost-based optimization of these workflows remains a challenge due to the multitude of interfaces, large size of the execution plan space, and the frequent unavailability of all types of information needed for optimization. We introduced Stubby as a comprehensive solution to this problem.

Stubby is an extensible, cost-based, and transformation-based workflow optimizer that works across different interfaces for generating MapReduce workflows. Stubby is designed to sit above the MapReduce system, but below and external to any software system that submits workflows to the MapReduce system. Depending on the information available, Stubby considers all valid transformations from the full plan space (which we described in detail) to cost and pick the near-optimal set of transformations to apply on an input workflow. A comprehensive experimental evaluation showed the effectiveness of Stubby which generated optimized workflows with speedups of up to 4.5X over the baseline.

## APPENDIX

An online technical report version of this paper is available at [12]. This report contains the following supplementary material:

- Proof of the conditions for intra-job vertical packing.
- Pig Latin queries of workflows used in the evaluation section.
- Details of how annotations are generated.